\documentclass[9pt,sigconf, nonacm]{acmart}

\renewcommand\footnotetextcopyrightpermission[1]{} % removes footnote with conference information

\usepackage{amsmath}
\usepackage{graphicx}
\usepackage{textcomp}
\usepackage{xcolor}
\usepackage{makecell}
\usepackage{multirow}
\usepackage{float}
\begin{document}
\title[]{Algorithm-Hardware Co-Design of Distribution-Aware Logarithmic-Posit Encodings for Efficient DNN Inference}
\titlenote{2024 61st IEEE/ACM Design Automation Conference (DAC)}
\author{Akshat Ramachandran$^1$, Zishen Wan$^1$, Geonhwa Jeong$^1$, John Gustafson$^2$, Tushar Krishna$^1$}
% \author{}
\affiliation{
  \institution{$^1$Georgia Institute of Technology, Atlanta, GA, $^2$Arizona State University, Tempe, AZ}
  \country{}
}

%inefficient low-precision encoding, , reduced compression rate, larger silicon footprint, and/or require computationally intensive quantization-aware training.

% LP
% DQF DistriBit Quantization Framework
% DistriBit PE
% LPNS vs LP
% Number system  --- LPNS / Floating point
% Data Type      --- LP   / Float16 / adaptive float

%%%%%%
% TODO: Read the whole paper and check grammar and overall flow
% Check all figures and any missed cross-references
%%%%%%%

\begin{abstract}
Traditional Deep Neural Network (DNN) quantization methods using integer, fixed-point, or floating-point data types struggle to capture diverse DNN parameter distributions at low precision, and often require large silicon overhead and intensive quantization-aware training.
In this study, we introduce Logarithmic Posits (LP), an adaptive, hardware-friendly data type inspired by posits that dynamically adapts to DNN weight/activation distributions by parameterizing LP bit fields. We also develop a novel genetic-algorithm based framework, LP Quantization (LPQ), to find optimal layer-wise LP parameters while reducing representational divergence between quantized and full-precision models through a novel global-local contrastive objective.
Additionally, we design a unified mixed-precision LP accelerator (LPA) architecture comprising of processing elements (PEs) incorporating LP in the computational datapath. Our algorithm-hardware co-design demonstrates on average <1\% drop in top-1 accuracy across various CNN and ViT models. It also achieves $\sim 2\times$ improvements in performance per unit area and $2.2\times$ gains in energy efficiency compared to state-of-the-art quantization accelerators using different data types. Code available at: \textcolor{cyan}{\url{https://github.com/georgia-tech-synergy-lab/LogarithmicPosit}}
\end{abstract}

% Include this for CCS Concepts
%\include{includes/ccs_concepts}
% \keywords{Computer Arithmetic, Quantization, Deep Learning, Computer Vision}

\maketitle
\pagestyle{empty} % removes running headers
\setlength{\belowdisplayskip}{0pt} \setlength{\belowdisplayshortskip}{0pt}
\setlength{\abovedisplayskip}{0pt} \setlength{\abovedisplayshortskip}{0pt}
\vspace{-5pt}\section{Introduction}
\begin{figure}[t]
    \centering\includegraphics[width=0.9\columnwidth, keepaspectratio]{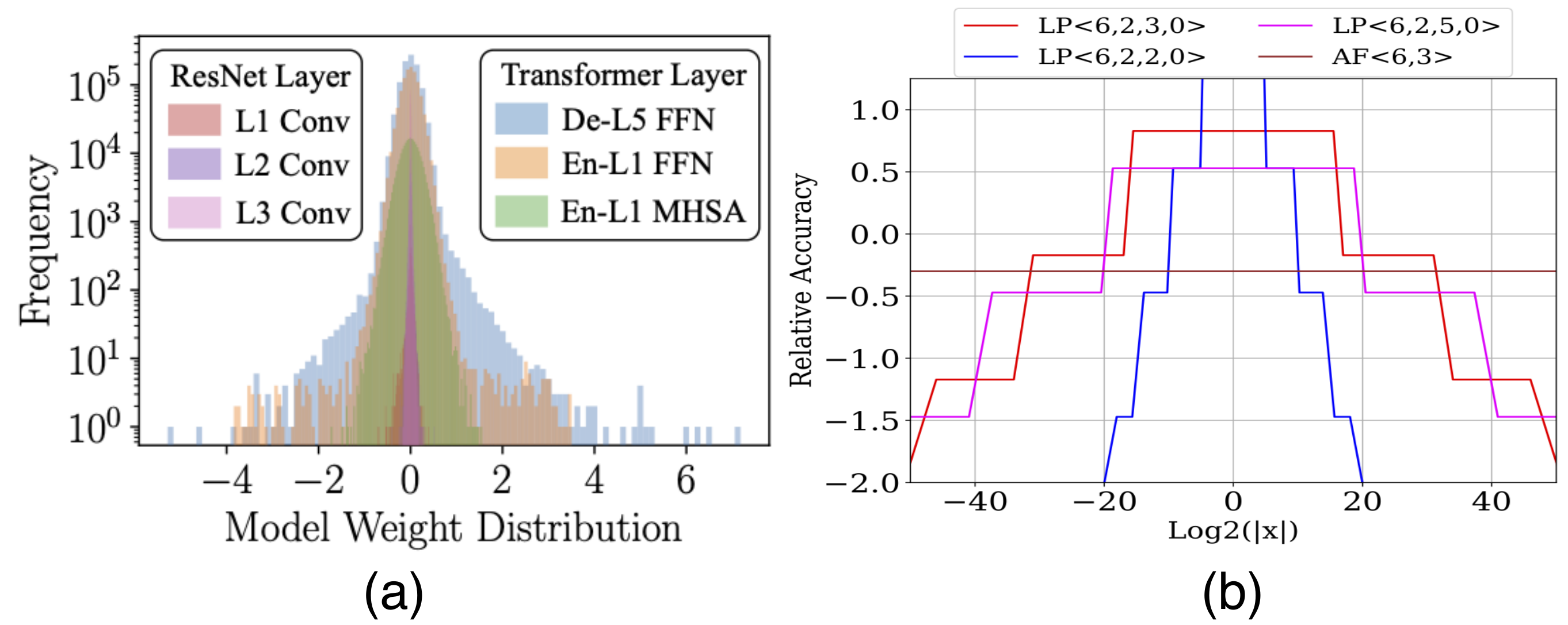}
        \vspace{-10pt}
        \caption{(a) Weight distributions of ResNet50 and ViT (De: Decoder, En: Encoder) layers, (b) LP's relative-accuracy plot, showing distribution-aware properties compared to AF \cite{tambe2020algorithm}.}
        \label{fig:weight_distribution}
        \vspace{-15pt}
\end{figure}
In response to the escalating computational and storage demands of DNNs, compressing models before deploying them on edge devices and cloud servers has become imperative \cite{tambe2020algorithm, liu2021improving}. Quantization has emerged as one of the most promising solutions to address the challenges of deploying DNNs on resource-constrained devices.

Along these lines, numerous techniques focus on uniform quantization, representing values as integers \cite{sharma2018bit} or fixed-point numbers. However, as Fig.\ref{fig:weight_distribution}(a) illustrates, there is substantial distributional variance and orders of magnitude differences in DNN parameters between layers and across models, leading to significant quantization errors when applied to modern DNNs \cite{guo2022ant}.

Seeking wider dynamic range and distribution-adaptive data formats, interest has grown in non-uniform quantization methods involving floating-point \cite{tambe2020algorithm, liu2021improving}, posits \cite{langroudi2019cheetah, ramachandran2022positiv}, and logarithmic number systems (LNS) \cite{alam2021low}. Adaptive floating-point techniques such as \cite{tambe2020algorithm} adjust the exponent range empirically based on the dynamic range of parameters. However, they fail to adapt to the tapered distribution of DNN parameters and use flat accuracy. Floating-point encodings also come with increased hardware complexity, wasted bit patterns, and convoluted exception handling, hindering adoption in edge devices \cite{murillo2020deep}.

Recently, posit-based representations have demonstrated advantages over standard floats for DNN inference, offering tapered accuracy (due to run-length encoded regime), providing a larger dynamic range, higher accuracy, and simpler exception handling \cite{gustafson2017beating, murillo2020deep}. Posit hardware, though cheaper than float hardware, is still not efficient enough for adoption in resource-constrained devices. Inspired by the efficiency of integers combined with the benefits of floats in \cite{guo2022ant}, we propose \textbf{Logarithmic Posits (LP)}, a composite data type that blends the adaptability of posits with the hardware efficiency of LNS. LP exploits the tapered accuracy of posits (regime), exponent size and scale factor (exponent bias) to tailor the representation range, shape and position to the DNN parameter distribution while capitalizing on the computational efficiency of LNS (Fig.\ref{fig:weight_distribution}(b)).

To utilize LP for DNN quantization, we introduce an automated \textbf{LP Quantization (LPQ) Framework} based on genetic algorithms. LPQ operates in a Post Training Quantization (PTQ) setting with access to a small unlabelled calibration dataset ($128$ images). Building on previous works \cite{cai2020zeroq, frumkin2023jumping, liu2021improving}, we incorporate a novel global-local contrastive objective to combat over-fitting to the calibration data and prevent premature convergence by minimizing divergence between intermediate representations (intermediate layer output activations) of the quantized and full precision (FP) model.

To efficiently execute computations with LP, we further propose a mixed-precision \textbf{LP Accelerator (LPA)} that integrates mixed-precision LP processing elements (PEs) into a systolic array architecture. Our co-design targets mixed-precision quantization of DNNs. Extensive experiments on CNNs and ViTs demonstrate <1\% drop in top-1 accuracy across model families and surpass state-of-the-art mixed-precision accelerators \cite{guo2022ant, sharma2018bit} with $~2\times$ performance per unit area improvement and $2.2\times$ energy reduction.
%%%%%%%%%%%

% \begin{figure}[t]
%     \centering\includegraphics[width=0.7\columnwidth, keepaspectratio]{figures/LogPosits_Regime_Dist.pdf}
%         \vspace{-10pt}
%         \caption{Relative accuracy plot of Logarithmic Posits showing the fine grained control achieved with the regime as compared to the flat accuracy achieved with Adaptive Floats \cite{}}
%         \label{fig:logpos_regime}
%         \vspace{-10pt}
% \end{figure}

\section{Background and related work}
\label{sec:bg}
\textbf{Standard Posit Representation.}The standard posit format \cite{gustafson2017beating} with size $n$-bits includes the sign, regime $(r)$ of size \emph{rs}, exponent $(e)$ of size \emph{es}, and fraction $(f)$ fields. Unlike floats, posits encode the regime field using a run-length $m$ of $0$s or $1$s terminated by a $1$ or $0$, respectively, or by the final bit. The regime value $k$ is determined as $k= -m$ if the first bit of the regime is $0$, or $k= m-1$ otherwise. 
The regime creates tapered accuracy in the posit representation (Fig.\ref{fig:weight_distribution}(b)) unlike floats, which has a flat accuracy.
This property, which is beneficially used in prior works such as \cite{langroudi2019cheetah, murillo2020deep}, is particularly useful. In PositNN \cite{langroudi2019positnn}, \emph{rs} is manually configured for DNNs, but this may not guarantee performance across different models. In contrast, we propose parameterizing all posit bit-fields to enhance adaptability without handcrafted tuning, while maintaining a hardware-friendly representation (Sections \ref{sec:lp} and \ref{sec:lpq}). Achieving similar adaptability in standard LNS would require a combination of arbitrary bases \cite{alam2021low}, leading to complicated arithmetic circuitry.

\textbf{Quantization Objective. }Previous studies leverage conventional loss functions like KL-Divergence, mean-squared-error (MSE), and cosine similarity as a global quantization objective (final output) to determine the best parameters \cite{liu2021improving, cai2020zeroq, guo2022ant}. In our PTQ framework, traditional loss functions tend to overfit to the calibration data and lack generalization to the test set.
Furthermore, relying solely on the final output for the quantization search process can lead to premature convergence or sub-optimal solutions as the search progresses, ignoring the representational collapse of intermediate layer outputs compared to the FP model. Contrastive loss functions, common in self-supervised settings, have been proven by prior work \cite{fradkin2022robustness} to combat overfitting by regularizing against negative samples in the test set. We leverage a global-local contrastive loss, estimating the representational divergence of intermediate representations (intermediate layer output activations) in addition to the final output to identify the best precision, preventing the representational collapse of the quantized model. (Section. \ref{sec:fitness})

\textbf{Quantization Accelerators. }There is a recent surge in DNN inference accelerators embracing mixed-precision techniques and novel data types. BitFusion features fusible low-precision integer PEs within a systolic-array architecture \cite{sharma2018bit}. AdaptivFloat introduces adaptive floating-point quantization and hybrid float PEs to mitigate integer quantization errors, albeit with substantial area overheads \cite{tambe2020algorithm}. Improving on prior works \cite{tambe2020algorithm, sharma2018bit}, the ANT \cite{guo2022ant} design employs a 4-bit INT PE with decoders to support float computations on the same INT PE. Recent efforts also explore benefits of posits over float for DNN inference, like the fixed-resolution posit MAC unit and the mixed-precision posit PE in \cite{langroudi2019cheetah}. Despite posits' superiority over floats, the high resource utilization in prior works hinders adoption in resource-constrained devices. Inspired by \cite{guo2022ant} and posit's advantages, in this work we propose a Logarithmic Posit PE design, that exploits both the higher accuracy and adaptability of posits (Fig.\ref{fig:weight_distribution}(b)) and the computational efficiency of LNS.
\section{LP: Logarithmic Posits}
\label{sec:lp}
The proposed Logarithmic Posit data type (LP) closely follows the general scheme of the standard posit format \cite{gustafson2017beating} while leveraging the computational/hardware efficiency of LNS \cite{alam2021low}. We parameterize several additional bit fields of a standard posit, which provides fine-grained control over the dynamic range, position, and shape of the distribution of number encodings to better emulate the heterogeneous weight/activation distributions of DNN encodings. The parameters we incorporate are:

\textbf{Number of bits $(n)$. } We dynamically adjust the number of bits to allow mixed-precision quantization of a DNN to enable us to choose the optimal precision for a layer and achieve higher compression rate.

\textbf{Exponent Size} (\emph{es}). Modifying \emph{es} allows LP to adapt to diverse dynamic ranges. Each increment in \emph{es} doubles the dynamic range. \emph{es} is limited to $n-3$ bits to allow 1-bit sign and atleast 2-bits regime.

\textbf{Regime Size} (\emph{rs}). This parameter enables us to control the degree of tapering of the number system (shape of distribution) as highlighted in Fig.\ref{fig:weight_distribution}(b). 
Standard posits have a fixed tapering for all precisions. $rs$ is constrained to at most $n-1$ bits to allow 1 bit for the sign.
The fraction field $(f)$ occupies the remaining bits, if any. Because there is an implied value before the radix point (similar to the hidden bit in floats), the absence of fraction bits does not represent zero.

\textbf{Scale Factor} \emph{sf}. The scale factor is a continuous-valued parameter that biases the scaling of the representation upwards or downwards. By adding a scale factor bias we can shift the region of maximum accuracy of posits (tapered region) to the desired region. In standard posits there is no scale factor bias and the tapered region is always centered at magnitude $0$.

Finally, inspired by the arithmetic efficiency and low hardware cost of LNS \cite{alam2021low}, we augment the parameterized posit representation above to include its advantages.  We express the standard posit fraction $(1.f)$ and exponent fields in the logarithmic domain as a unified fixed-point exponent of the power of two as $2^{e + f'}$, where $f' = log_2(1.f)$. $f'$ corresponds to the fractional part and $e$ to the integer part of the fixed-point format which we term Unified Logarithmic Fraction and Exponent (\textbf{ulfx}). 

Mathematically, LP can be represented as,
\begin{equation}
    x \langle n, \textit{es}, \textit{rs}, \textit{sf}\rangle = (-1)^\textit{sign}\times2^{2^\textit{es}\times k- \textit{sf}}\times 2^{\mathbf{ulfx}}
\end{equation}
\section{LPQ: LP Quantization Framework}
\label{sec:lpq}

\begin{figure}[t] \centering
\includegraphics[width=\linewidth]{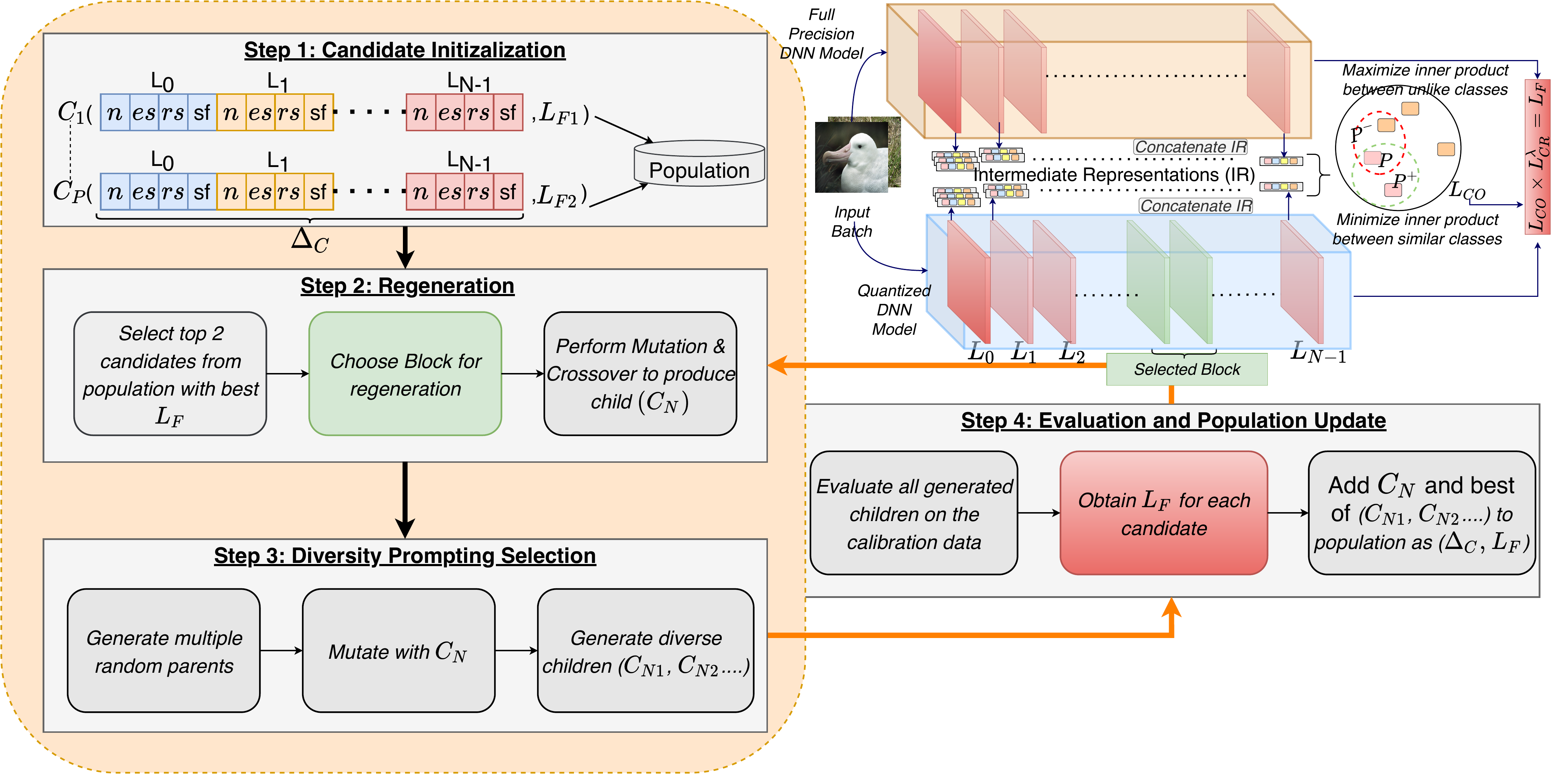}
\vspace{-20pt}
\caption{Overview of LPQ Framework illustrating the four major steps and evaluation of fitness function.}
\vspace{-20pt}
\label{fig:overview}
\end{figure}

We present an overview of our LPQ framework in Fig. \ref{fig:overview}, which is composed of four stages.

\textbf{\textit{Step 1: Candidate Initialization. }}A quantization solution comprises an encoded vector $\Delta$ of length $4N$ and each set of $4$ values represent the $4$ LP parameters of a layer $l$: $\Delta[l] = \langle n_l, es_l, rs_l, sf_l \rangle$. We constrain the search space as follows,  $n$ within $[2,8]$, \emph{es} and \emph{rs} within $[0, n-3]$ and $[2, n-1]$. Following previous work \cite{frumkin2023jumping}, which highlights quantization sensitivity to small scale perturbations, we extend this to \emph{sf}. The \emph{sf} search space for each layer $l$ is a uniform ball of radius $10^{-3}$ centered around the mean weight distribution of that layer. Prospective scale factors are sampled as $\textit{sf}^{\;l} = \text{mean}(l) + \eta(-10^{-3}, +10^{+3})$, where $\eta$ is a random sampling function. LPQ initiates the population by randomly sampling $\mathcal{K}$ candidate $\Delta$ vectors consisting of different quantization strategies per layer. The fitness function $\mathcal{L}_F$ is evaluated for each candidate (used to identify best candidates in later stages, explained in Sec.~\ref{sec:fitness}). We create $\mathcal{K}$ tuples $(\Delta_k, \mathcal{L}_F^k)$ to form the initial population. Fitness values of initial candidates are pre-computed and stored to avoid repeated evaluations.

\textbf{\textit{Step 2: Re-generation (Crossover and Mutation).}} Each candidate in the population is ranked based on the fitness function and the top two candidates serve as parents for generating the next candidate generation (child). When evolving candidates, perturbing too many layer parameters based on parents can lead to a high-dimensional search space; to mitigate this, we employ a block-wise regeneration approach, evolving only a subset/block of size $\mathcal{B}$ of child parameters based on chosen parents, setting all others to the best parent's parameters. The child's parameter regeneration for the specific block involves adjustments according to parent candidates ($p1$, $p2$) and is formulated as:
    \begin{gather}
        n_{\textit{child}} = \mathbf{rand}(\mathbf{min}(p1.n, p2.n)-1, \mathbf{max}(p1.n, p2.n) + 1) \\
        es_{\textit{child}} = \mathbf{rand}(\mathbf{min}(\text{p1}.\textit{es}, p2.es)-1, \mathbf{max}(p1.es, p2.es) + 1) \\
        rs_{\textit{child}} = \mathbf{rand}(0, \mathbf{ceil}(\mathbf{mean}(p1.rs, p2.rs))+1) \\
        sf_{\textit{child}} = \mathbf{mean}(p1.sf, p2.sf) + \eta(-10^{-3}, 10^3)
    \end{gather}
We prefer $\mathbf{mean}()$ for parameters that influence the shape of the distribution of the number encodings and $\mathbf{min}()/\mathbf{max}()$ for parameters that affect the dynamic range.

\textbf{\textit{Step 3: Diversity Promoting Selection. }}Instead of directly adding the regenerated child back into the population for use in the next iteration of the search process, we propose to introduce diversity into the population and prevent premature convergence. To this end, we create additional random parents (empirically chosen to be five in this work) and use the regenerated child in the previous stage as the other parent to generate five diverse children.

\textbf{\textit{Step 4: Evaluation and Population Update. }} We evaluate all generated children in Step 2 and 3 and acquire the fitness function. The child generated in Step 2 and the corresponding fitness function value is added to the population. We then rank the diversity promoting children in Step 3 and select the best child to be added to the population for the next iteration.

In our block-wise genetic algorithm search strategy, we employ $\mathcal{P}$ passes over the whole DNN, i.e. over all the blocks of size $\mathcal{B}$, and each block is iterated over $\mathcal{C}$ cycles in each pass. Therefore, the population is updated $\mathcal{P} \times \mathcal{C}$ times , i.e., Steps 2, 3, and 4 are executed $\mathcal{P} \times \mathcal{C}$ times. 

\textbf{\textit{Quantization for Activation.}} After determining the quantization parameters for all DNN weights, we identify the LP quantization values for each input activation in the corresponding layer. Activation quantization sensitivity closely aligns with that of the weight parameters producing them. The LP parameters of output activation of layer $l$ are $n_{act}^l = \mathbf{min}(8, n_w^l \times 2)$, $es_{act}^l = \mathbf{min}(5, es_w^l \times 2)$, and $\textit{sf}_{act}^{\;l} = \textit{sf}_{act}^{\;l-1} + \textit{sf}_w^{\;l}$. We find that retaining the regime, i.e. $rs_\textit{act}^l = \textit{rs}_w^{\;l}$,  achieves best performance.   

\vspace{-2mm}
\subsection{Fitness Function}
\label{sec:fitness}
We introduce a fitness function, $\mathcal{L}_F$, to evaluate quantization strategies. It assesses two key metrics: intermediate representation divergence and compression ratio relative to the FP model. The representation divergence metric aims to align the distribution of quantized model's intermediate representations closely with the FP model, while the compression ratio metric incentivizes lower bit-widths.

We formulate a combined global-local contrastive objective to address limitations in traditional loss functions. Let the concatenated tensor of intermediate representations from each layer of the FP and quantized model be denoted as $\textbf{H}^\text{FP} = \{\textbf{H}_0^\text{FP}, \textbf{H}_1^\text{FP}, \ldots, \textbf{H}_{N-1}^\text{FP} \}$ and $\textbf{H}^{q} = \{\textbf{H}_0^{q}, \textbf{H}_1^{q}, \ldots, \textbf{H}_{N-1}^{q} \}$, where $\textbf{H}_l$ represents the intermediate output tensor after passing through layer $l$. However, using $\textbf{H}^\text{FP}, \textbf{H}^{q}$ directly is impractical due to high dimensionality. We address this by applying row-wise pooling using \textit{Kurtosis-3} \cite{kurtosis_3} instead of mean pooling. \textit{Kurtosis-3} better characterizes distribution tailedness of DNN parameters. The contrastive objective for representational divergence estimation is thus formulated in Equation. \ref{contrastive_loss}, closely following the contrastive loss definition in \cite{fradkin2022robustness}.

\begin{equation}
\label{contrastive_loss}
    \mathcal{L}_\text{CO} = \log(1+e^{-\langle\mathcal{H}^q_p, \mathcal{H}^\text{FP}_{p+}\rangle/ \tau}
                                \sum_{p-} e^{\langle\mathcal{H}^q_p, \mathcal{H}^\text{FP}_{p-}\rangle/ \tau})
\end{equation}
\\
\noindent
where $\tau$ controls concentration level \cite{fradkin2022robustness} following the typical definition used in contrastive loss literature; $p$, $p+$, and $p-$ are quantized model prediction on a particular image, the corresponding FP model prediction (positive example), and FP model predictions for all other calibration data images (negative examples).

We further penalize higher bit-width candidates using a loss that quantizes the compression factor inspired by \cite{liu2021improving} as, $\mathcal{L}_\text{CR} = \sum_{l \in N} \#\text{PARAM}(\mathbf{H}_l ^ \text{FP}) \times n_l$. The complete fitness function $\mathcal{L}_\text{F}$ is defined as a combination of the two components defined above, balanced by a coefficient $\lambda$, $\mathcal{L}_\text{F} = \mathcal{L}_\text{CO} \cdot \mathcal{L}_\text{CR}^\lambda$. In our experiments, we empirically set the parameter $\lambda=0.4$ to achieve the best compromise between representational accuracy and compression factor.
\section{LPA: LP-based DNN accelerator}
In this section, we introduce the modifications to a systolic array architecture to support LP and the design of an LP PE to natively handle multiple precisions and parameter sets.

\subsection{Architecture Overview}
\label{acc_arch}
\begin{figure}[t]
% \vspace{-2pt}
    \centering\includegraphics[width=\columnwidth]{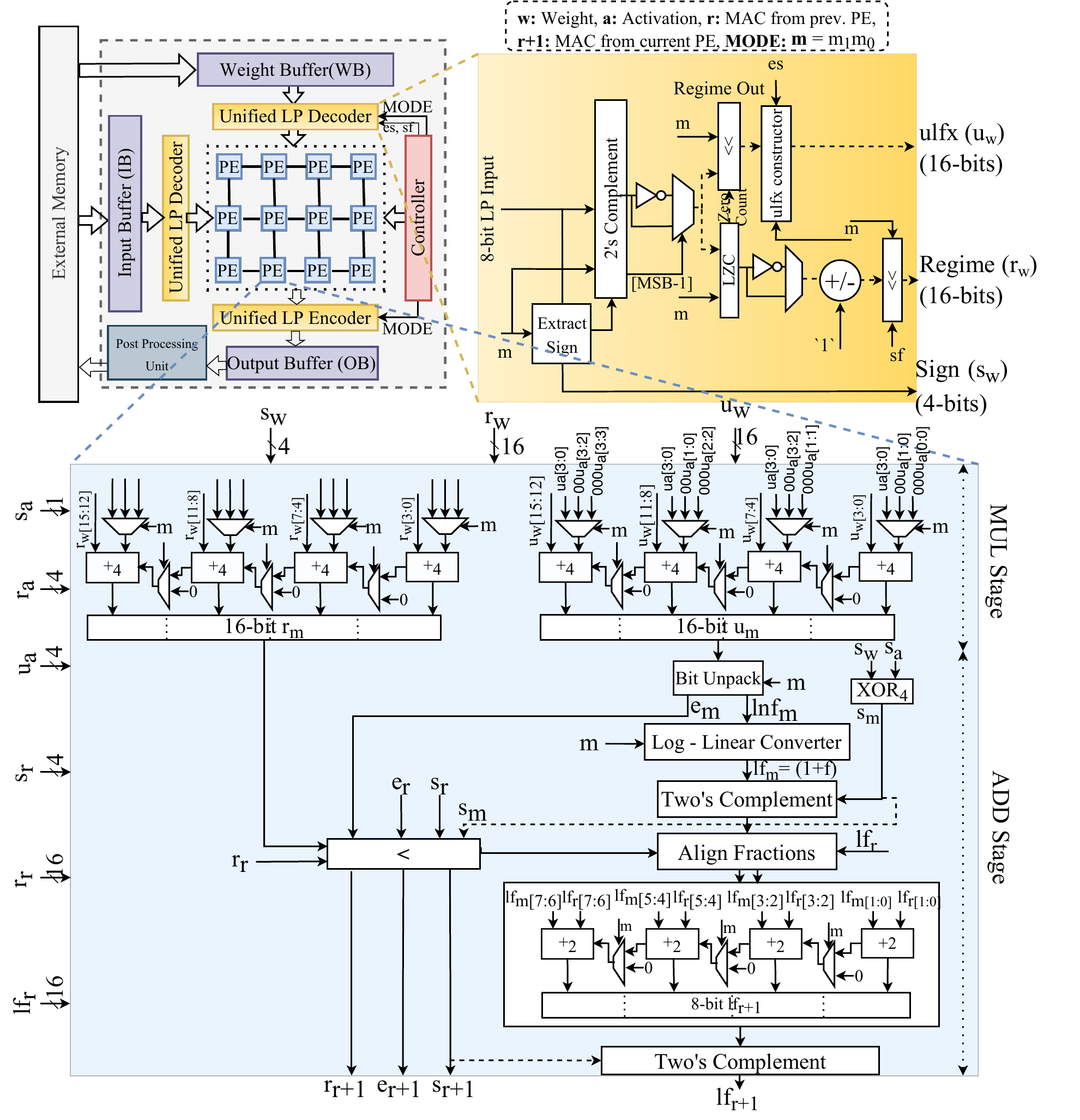}
        \vspace{-20pt}
        \caption{LPA Architecture depicting detailed LP PE and Unified LP Decoder units.}
        \label{fig:lpa}
        \vspace{-20pt}
\end{figure}

Fig. \ref{fig:lpa}, shows LPA with architectural optimizations for LP support on a systolic array. Our design optimizes computation throughput in a weight-stationary dataflow, enabling the mapping of multiple low-precision weights sharing an eastbound input activation to a single PE.

\textbf{Weight and Activation Organization. } For hardware efficiency and bit packing during DNN inference on LPA, we constrain the LPQ search space of $n$ to integer powers of 2. For weights, $n$ ranges from 2 to 8, and corresponding activations are 4- or 8-bit. The weights and input activations are stored as 8-bit LP in their corresponding buffers. In the input buffer, the 4-bit activations are also stored as an 8-bit value by zero-extending the LSB. The interpretation of the weight buffer's bit pattern depends on the quantization precision of the mapped weights and is identified by the \textbf{MODE} (provided by the controller). Each PE supports three modes based on the quantization precision: MODE-A (four 2-bit weights), MODE-B (two 4-bit weights), and MODE-C (one 8-bit weight).

\begin{figure}[t]
\vspace{-5pt}
    \centering\includegraphics[width=0.88\columnwidth]{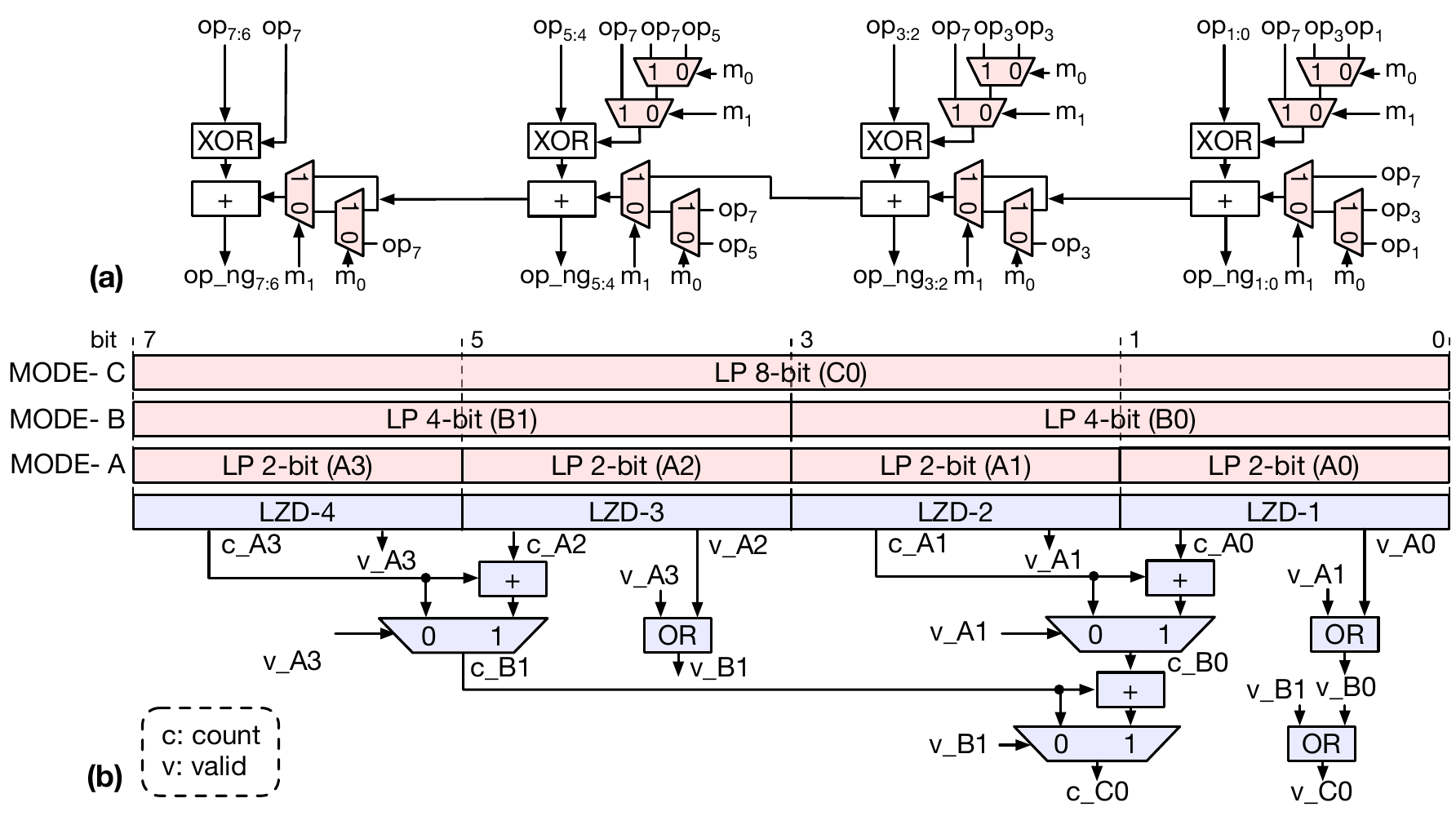}
        \vspace{-10pt}
        \caption{Architecture of mixed-precision 2's complementer (a) and Leading Zero Detector (LZD) (b).}
        \label{fig:pe_decoding}
        \vspace{-15pt}
\end{figure}

% \TK{I think this para should be shortened to save space .. }
\textbf{LP Decoder. } 
We insert LP decoders between on-chip buffers and the PE array, strategically placing them only along the boundary to minimize area overhead. Quantized tensors are stored in low precision both on-chip and off-chip, and LP decoders are employed to convert them to a unified format. For an 8-bit LP weight in the Weight Buffer (WB), we decompose it into sign (4 bits), regime (16 bits), and \textbf{ulfx} (16 bits), as illustrated in Fig. \ref{fig:lpa}. The decoding process begins with a unified 2's complementer, highlighted in Fig. \ref{fig:pe_decoding}(a) that handles multiple precisions simultaneously controlled by multiplexers. The regime is then decoded by counting the number of leading ones or zeros after the sign bit. To avoid the implementation of both a leading zero and a leading one counter, the binary is inverted according to the regime’s first bit. The LZC (Fig. \ref{fig:pe_decoding}(b)), similar to the unified 2's complementer, provides the zero count of multiple precision inputs. Based on the zero count, the regime value is shifted out from each LP using four small left shifters. Depending on the MODE, the shifted value from one shifter may be sent to the next shifter. 

After shifting out the regime, the remaining bits form the \textbf{ulfx}, consisting of the exponent and log-domain fraction. The \textbf{ulfx} is interpreted as a fixed-point number with equal bit allocations to the exponent (integer part) and log-domain fraction (fraction part). For example, in MODE-B, the \textbf{ulfx} is interpreted as two fixed-point numbers, each with 4-bit integer part and 4-bit fraction part. The zero count (in 2's complement) is also used to calculate the regime value, adjusted for scale factor, and stored as a 16-bit regime value. The higher precision to represent \textbf{ulfx} and regime in the unified format is chosen to prevent overflow during calculations. The activation decoder follows a similar process, except the outputs are a 1-bit sign, 4-bit regime, and 4-bit \textbf{ulfx}. These sizes facilitate easier routing to the adders in the Processing Element (PE).

\textbf{LP Encoder. } The unified LP encoder, mirroring decoder primitive components, performs the inverse operation, packing LP components into an 8-bit zero-extended format. The encoder is also responsible for converting the linear domain (lf) fractions of the partial sums into the log-domain (lnf) (explained later).

\textbf{Post Processing Unit (PPU). } The PPU is configured based on the controller to quantize the partial-sum outputs from the PE-array to either 4- or 8-bit LP, calculate activation scale factors, and perform non-linear operations (ReLU/Softmax) similar to \cite{keller202395, gustafson2017beating}.

%%TK - i moved this to next subsection
%\textbf{PE Array Organization. }Unlike previous mixed-precision systolic array architectures such as \cite{guo2022ant, sharma2018bit}, which utilize low-precision PEs and combine multiple PEs to support higher precisions, our approach is inspired by recent work \cite{jeong2023vegeta}. We propose mapping multiple weights (MODE-A/-B/-C) sharing the same input activation to a single PE. With lower precision weights in a layer, more weights can be assigned to the same PE, enabling parallel evaluation of more partial-sums along the same PE column. This amortizes the overhead of memory, control, and horizontal activation transfer across a row of the array, improving computation throughput.

\subsection{PE Architecture}
The weight-stationary PE in the array is double-buffered with the ability to store decoded weights for the next computation, allowing amortization of decoding and filling the systolic-array PEs for each computation. Each PE receives decoded activations from the left and partial sums from the top which are propagated to the right and bottom respectively, every cycle.

\textbf{Multi-precision.} In contrast to earlier mixed-precision systolic array architectures like \cite{guo2022ant, sharma2018bit}, which employ low-precision Processing Elements (PEs) and combine multiple PEs to support higher precisions. We suggest a strategy of mapping multiple weights (depends on MODE-A/-B/-C) that share the same input activation to a single PE. By using lower precision weights in a layer, it becomes possible to assign more weights to the same PE. This facilitates the parallel evaluation of multiple partial sums along the same PE column, thereby amortizing the overhead of memory and control, and increasing performance per unit area.

\textbf{Multiplication Stage. } In LP, the multiplication of weights and activations is replaced by addition of \textbf{ulfx} and regimes. As shown in Fig.\ref{fig:lpa}, this is done in parallel using 2-sets of four 4-bit adders. The activation's regime and \textbf{ulfx} to be added are chosen based on the MODE. In MODE-A, each weight has to be added with each activation, therefore the complete \textbf{ulfx} and regime are passed to the corresponding adders with no carry propagation between adders. Similarly, in MODE-B/-C, the activation's \textbf{ulfx} and regime bits are split into multiple lower bit-width components, zero-extended and passed to each of the weights individually. The MUL-stage results in a 16-bit regime, 16-bit \textbf{ulfx} and is guaranteed to not overflow. The 16-bit \textbf{ulfx} is split into 8-bit exponent and 8-bit fraction (lnf) for the next stage.

\begingroup	
\begin{table*}[t!]\centering
\vspace{-5pt}
 \caption{Quantization accuracy comparison against competing methods on ResNet18, ResNet50 and MobileNetV2.}
\vspace{-10pt}
\resizebox{0.8\linewidth}{!}{%
\begin{tabular}{c|ccc|ccc|ccc}
\Xhline{2\arrayrulewidth}
  & \multicolumn{3}{c|}{ResNet18} & \multicolumn{3}{c|}{ResNet50} & \multicolumn{3}{c}{MobileNetV2} \\
\Xhline{1\arrayrulewidth}
Method & W/A & Model Size(MB) &  Top-1 Accuracy & W/A & Model Size(MB) &  Top-1 Accuracy & W/A & Model Size(MB) &  Top-1 Accuracy \\
\Xhline{2\arrayrulewidth}
Baseline & 32/32 & 44.60 & 71.08 & 32/32 & 97.80 & 77.72 & 32/32 & 13.40  & 72.49 \\
\Xhline{1\arrayrulewidth}
EMQ \cite{dong2023emq} & MP/4 & 5.50 & 70.12 & MP/5 & 17.86 & 76.70 & MP/8 & 1.50 & 70.75 \\
HAWQ-V3 \cite{yao2021hawq} & 4/4 & 5.81 & 68.45 & MP/MP & 18.70 & 75.39 & MP/MP & 1.68 & 70.84 \\
AFP \cite{liu2021improving} & - & - & - & MP$_{4.8}$/MP & 13.20 & 76.09 & MP$_{4.8}$/MP & 1.94 & 70.91 \\
ANT \cite{guo2022ant} & MP/MP & 5.87  & \textbf{70.30} & MP/MP & 14.54 & 76.70  & MP/MP & 1.84 & 70.74  \\
BREC-Q \cite{li2021brecq} & MP/8 & 5.10 & 68.88 & MP/8 & \textbf{13.15} & 76.45 & MP/8 & \textbf{1.30} & 68.99 \\
% Q-PIM \cite{long2020q} & MP$_{5.1}$/MP & 5.20 & 68.90 & MP$_{5.2}$/MP & 13.70 & 74.60 &- &- &- \\
\textbf{LPQ (Ours)} & MP$_{4.2}$/MP$_{5.5}$ & \textbf{4.10} & \textbf{70.30} & MP$_{5.3}$/MP$_{5.9}$ & 14.0 & \textbf{76.98} & MP$_{4.1}$/MP$_{4.98}$ & \textbf{1.30} & \textbf{71.20} \\
\Xhline{2\arrayrulewidth}
\end{tabular}
}
\vspace{-10pt}
\label{table_cnn_accuracy}
\end{table*}
\endgroup

\begingroup	
\begin{table}[t]\centering
 \caption{Quantization accuracy comparison against SoTA methods on Vision Transformers (ViT-B, DeiT-S and Swin-T).}
 \vspace{-10pt}
  \renewcommand*{\arraystretch}{1.0}
  \setlength\tabcolsep{1.9pt}
\resizebox{\linewidth}{!}{%
\begin{tabular}{c|cc|cc|cc}
\Xhline{2\arrayrulewidth}
  & \multicolumn{2}{c|}{ViT-B} & \multicolumn{2}{c|}{DeiT-S} & \multicolumn{2}{c}{Swin-T} \\
\Xhline{1\arrayrulewidth}
Method & W/A &  Top-1 Accuracy & W/A &  Top-1 Accuracy & W/A &  Top-1 Accuracy \\
\Xhline{2\arrayrulewidth}
Baseline & 32/32  & 84.53 & 32/32  & 79.80 & 32/32  & 81.20 \\
\Xhline{1\arrayrulewidth}
Evol-Q \cite{frumkin2023jumping} & 4/8 & 79.50 & 4/8  & 77.06 & 4/8  & 80.43 \\
FQ-ViT \cite{lin2021fq} & 4/8 & 78.73 & 4/8 & 76.93 & 4/8 & 80.73 \\
\textbf{LPQ (Ours)} & MP$_{4.7}$/MP$_{6.3}$ & \textbf{80.14} & MP$_{3.9}$/MP$_{5.5}$ & \textbf{78.01} & MP$_{4.5}$/MP$_{6.2}$ & \textbf{80.98} \\
\Xhline{2\arrayrulewidth}
\end{tabular}
}
\vspace{-15pt}
\label{table_vit_accuracy}
\end{table}
\endgroup
\textbf{Accumulation Stage. } This stage receives the split exponent, lnf and the 16-bit regime from the MUL-stage along with the exponent, lf (linear-domain fraction), regime and sign of the partial sum from the previous PE in the same column. While multiplication in the log-domain is cheap, addition is inefficient. Therefore, we convert lnf to lf (fraction in the linear domain i.e., $1.f$). Instead of implementing an expensive LUT based converter, inspired by \cite{alam2021low}, we use an 8-bit Log-Linear converter using a set of gates. The logic function for the converter is identified by using a Karnaugh-map solver on the truth table constructed for all possible log-linear conversions and all possible bit-pattern interpretations. The linear-log converter in the encoder is also implemented in a similar fashion but with an inverse truth table. After lf is obtained, it is two's complemented through a unified two's complementer, and a simple floating-point fraction alignment and scale-factor shifter logic is employed. The aligned fractions are added through four 2-bit adders to obtain the accumulated lf in 2's complement form along with the joint regime, 8-bit exponent, and 4-bit sign. The fraction is retained in the linear domain and not juxtaposed with exponent to prevent redundant conversions to linear domain since the partial sum output of a PE is always progressively accumulated. This is why the encoder employs a linear-log converter.
\section{Evaluation}
In this section, we evaluate the three contributions of the paper (LP, LPQ and LPA) on the aspects of quantization accuracy, performance, area, throughput, and energy efficiency. 

\textbf{Benchmarks and Datasets.} Our experiments are conducted on the ImageNet (ILSVRC2012) dataset, evaluating top-1 accuracy across various CNNs (ResNet18, ResNet50, MobileNetV2) and Transformer-based models (ViT-B, DeiT-S, and Swin-T) for computer vision tasks. The FP pre-trained models from pytorchcv serve as the baseline for our experiments.

We implement LPQ in PyTorch and employ a calibration dataset comprising $128$ randomly sampled images from the ImageNet training set. The algorithm's search parameters are empirically determined: Population Size ($\mathcal{K}$) = 20, Number of Passes ($\mathcal{P}$) = 10, Number of Cycles ($\mathcal{C}$) = 4, and Block Size ($\mathcal{B}$) is set to 4 for CNNs and one attention block for Transformer-based models.

LPA, consisting of LP PEs, decoders, and encoders, is implemented in Verilog RTL and synthesized using Synopsys Design Compiler with a TSMC 28 nm process. LPA is compared against three state-of-the-art baselines (refer Section.\ref{sec:bg}),  ANT \cite{guo2022ant}, BitFusion \cite{sharma2018bit}, and AdaptivFloat \cite{tambe2020algorithm}. For end-to-end performance evaluation of LPA and all baselines, we develop a cycle-accurate simulator tool based on DnnWeaver \cite{sharma2016high}. DeepScale \cite{sarangi2021deepscaletool} is employed to scale all designs to the 28 nm process for a fair comparison.

\subsection{Effectiveness of LPQ}
\begin{figure}[t]
    \centering\includegraphics[width=\columnwidth, keepaspectratio]{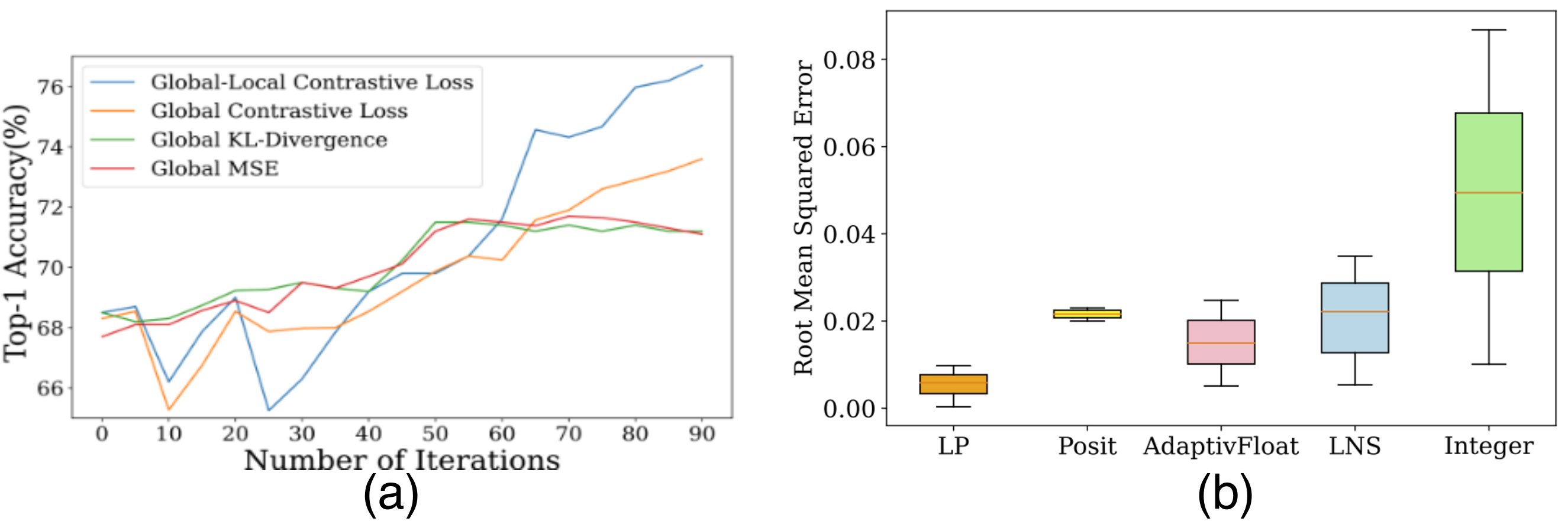}
        \vspace{-20pt}
        \caption{(a) LPQ performance with various loss functions, (b) RMSE distribution of quantization error of different formats.}
        \label{fig:eval_figs}
        \vspace{-10pt}
\end{figure}

\textbf{Number Format Comparison.} LPQ employs a novel data type, LP, consisting of two primitive data types—LNS and posits. We assess the impact of LP on quantization accuracy compared to its primitives and other conventional representations. LPQ is utilized for quantization of all data types, with modified search parameters suited to each data type for a fair comparison. Figure \ref{fig:eval_figs}(b) illustrates per-layer quantization error, measured with Root Mean Squared Error (RMSE), for various data types on ViT-B. LP consistently exhibits the lowest average RMSE, outperforming all other number formats. AdaptivFloat fares poorly compared to LP, primarily due to its limited ability to adapt only the dynamic range, lacking the distributional adjustment offered by LP.

\begingroup	
% \setlength{\tabcolsep}{3 pt}
% \vspace{-15pt}
\begin{table}[t]\centering
 \caption{Comparison of LPA with baselines under 28nm process with the same on-chip buffer (512kB (4.2 $mm^2$)).
 }
 \vspace{-10pt}
 \renewcommand*{\arraystretch}{1.0}
  \setlength\tabcolsep{1.9pt}
\resizebox{ \linewidth}{!}{%
\begin{tabular}{c|c|c|c|c|c}
\Xhline{2\arrayrulewidth}
 Architecture & Component (Area) & \makecell{Compute Area \\ ($\mu m^2$)}& \makecell{Throughput \\ (GOPS)} & \makecell{Compute Density \\ (TOPS/$mm^2$)} & \makecell{Total Area \\ ($mm^2$)}  \\
  \Xhline{2\arrayrulewidth}
 \multirow{3}{*}{LPA} & Decoder(5.2 $\mu m^2$)  & \multirow{3}{*}{12078.72} & \multirow{3}{*}{\textbf{203.4}} & \multirow{3}{*}{\textbf{16.84}}   & \multirow{3}{*}{4.212} \\ \cline{2-2}
                                  & Encoder (9.4 $\mu m^2$)& & & \\ \cline{2-2}
                                  & 2/4/8-bit PE (187.43 $\mu m^2$)& & &  \\ \cline{1-6}

 \multirow{2}{*}{ANT} & Decoder(4.9 $\mu m^2)$ & \multirow{2}{*}{5102.28} & \multirow{2}{*}{44.95} & \multirow{2}{*}{8.81} & \multirow{2}{*}{\textbf{4.205}}  \\ \cline{2-2}
                                  & 4/8-bit Int PE (79.57 $\mu m^2$) & & &  \\ \cline{1-6}

BitFusion & 2/4/8-bit PE & \textbf{5093.75} & 44.01 &  8.64 & \textbf{4.205}  \\ \cline{1-6}

AdaptivFloat & 8-bit PE & 23357.14 & 63.99 & 2.74 & 4.223   \\
\Xhline{2\arrayrulewidth}
\end{tabular}
}
\label{table_accelerator_baseline}
\vspace{-10pt}
\end{table}
\endgroup

\textbf{Comparison with State-of-The-Art (SoTA).} Our mixed precision quantization framework, LPQ, is compared against various competing works, both mixed-precision and uniform. Results are tabulated in Table \ref{table_cnn_accuracy} for CNNs and Table \ref{table_vit_accuracy} for ViTs. LPQ consistently outperforms other techniques, demonstrating <1\% average drop in accuracy. Notably, LPQ achieves lower average bit-widths for both weights and activations, resulting in an average compression of $7.5\times$. These outcomes can be attributed to two key factors: 1) LP's dynamic adaptation to the DNN parameter distribution, allowing for lower bit-width tolerance, and 2) the proposed fitness function components, which prevent representation collapse (contrastive objective) while encouraging lower bit-width (cost function).

\textbf{Convergence Behavior.} To validate the effectiveness of the proposed global-local contrastive loss, we compare it against common global loss functions—mean squared error (MSE), KL-divergence, and global contrastive loss \cite{frumkin2023jumping}. In Fig. \ref{fig:eval_figs}(a) %\ref{fig:loss_plot}
we observe that with increasing LPQ iterations, MSE and KL-Divergence curves plateau, indicating overfitting to the calibration dataset. Conversely, the global contrastive loss initially matches the performance of the global-local contrastive objective. However, as the number of iterations increases, the accuracy-gap widens because, the global contrastive objective fails to account for the representational collapse of intermediate representations as more layers undergo quantization.
%\vspace{-10pt}
\subsection{Effectiveness of LPA}
\textbf{Area.} We compare the accelerator area breakdown of LPA with the baselines in Table \ref{table_accelerator_baseline}. All accelerators have identical configurations, featuring an $8\times8$ weight stationary systolic array with same on-chip buffer configuration. The reported decoder and encoder area represents a single block for each row/column of the systolic array. The AdaptivFloat architecture, not supporting mixed-precision and limited to 8-bit \cite{tambe2020algorithm}, exhibits significantly larger area utilization due to its floating-point format. LPA PEs natively support 2/4/8-bit mixed precision. Whereas, ANT and BitFusion support 4-bit and 2-bit PEs respectively, achieving mixed-precision support by grouping neighboring PEs. Despite ANT and BitFusion exhibiting lower area when compared with LPA for the same number of PEs, LPA results in proportionately higher performance per unit area (TOPS/$mm^2$) for mixed-precision DNN inference.

\begingroup	
\begin{table}[t]\centering
 \caption{Impact on performance, accuracy and energy efficiency with different PE types in LPA.}
 \vspace{-10pt}
\resizebox{.9\linewidth}{!}{%
\begin{tabular}{c|ccc}
\Xhline{2\arrayrulewidth}
 PE-type & Density (TOPS/$mm^2$)  & Top-1 Accuracy &  Efficiency (GOPS/W) \\
  \Xhline{2\arrayrulewidth}
 LPA-2/4/8 & 16.84 & 76.98 & 212.17  \\
 LPA-8 & 6.98 & \textbf{77.70} & 124.26 \\
 LPA-2 & \textbf{23.79} & 0.0 & \textbf{438.96} \\
 Posit-2/4/8 & 3.15  & 73.65 & 70.36  \\
 AdaptivFloat-8 & 2.74 & 76.13 & 71.12 \\
 \Xhline{2\arrayrulewidth}
 
\end{tabular}
}
\vspace{-10pt}
\label{table_pe_types_accuracy}
\end{table}
\endgroup
\textbf{Performance Per Unit Area (TOPS/$mm^2$).} Using ResNet50 as the workload, we determine per-layer quantization parameters for LPA and BitFusion using LPQ. For ANT and AdaptivFloat, we adhere to the frameworks in their original papers. We ensure all baselines use quantization parameters that showcase their best possible accuracy for a fair comparison. In Table \ref{table_accelerator_baseline} (column 5), we present the performance per unit area of each design during quantized ResNet50 inference. LPA achieves nearly a $2\times$ improvement in performance per unit area compared to ANT and BitFusion for the same architecture configuration. Because these architectures tend to behave as 8-by-4 or 8-by-2 systolic arrays at higher precisions (because of PE fusion), LPA's advantage becomes pronounced by still maintaining an 8-by-8 behavior. To match LPA's performance, ANT/BitFusion would need wider systolic arrays, $8\times16$ or $8\times24$ respectively, offsetting their area advantage.

\begin{figure}[t]
    \centering\includegraphics[width=.85\columnwidth, keepaspectratio]{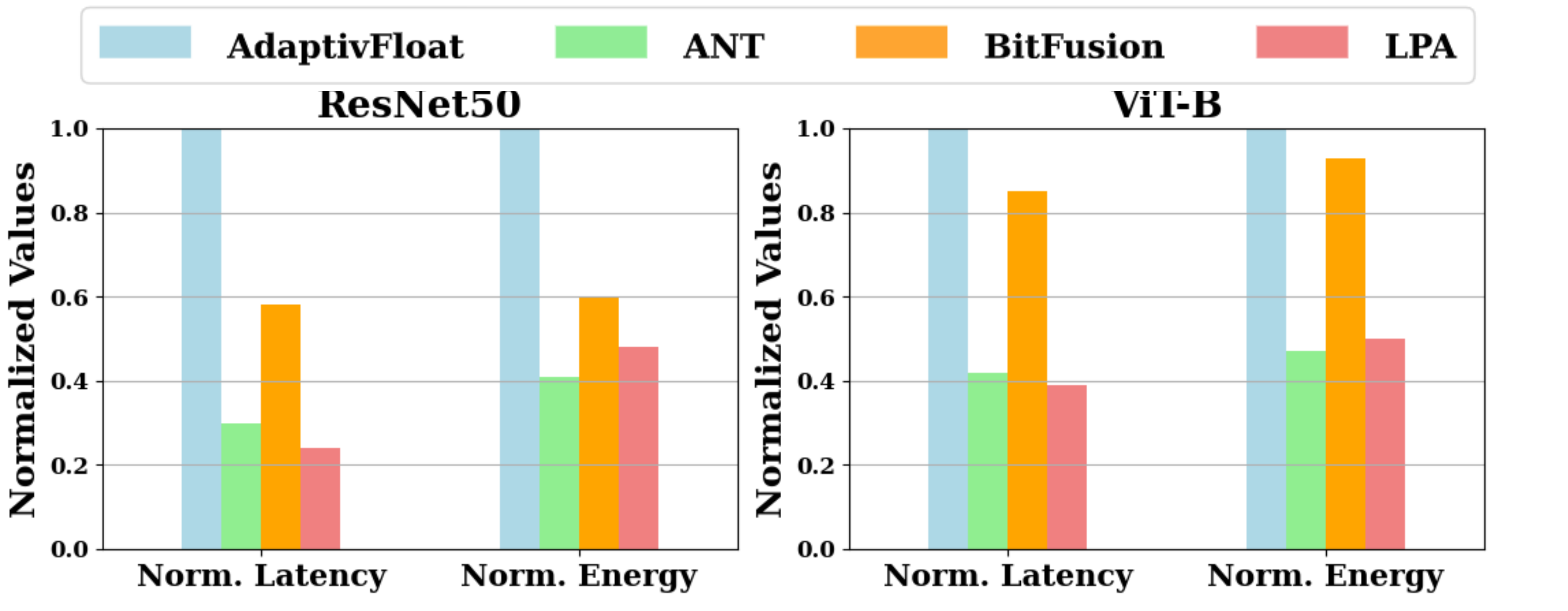}
        \vspace{-10pt}
        \caption{Normalized Latency and Energy comparison of LPA.}
        \label{fig:ppa_sota}
        \vspace{-12pt}
\end{figure}

\textbf{Performance and Energy Comparison with Baselines.} We compare LPA with the baselines on ViT-B and ResNet50, and report the normalized execution time and energy in Fig.~\ref{fig:ppa_sota}. LPA exhibits the lowest latency across models, with a modest increase in energy consumption over ANT attributed to overheads due to native mixed-precision support and conversion logic.

\textbf{Number Format and Mixed-Precision Comparison.} Examining the impact on performance, accuracy, and energy efficiency with different PEs supporting single-/mixed-precision for ResNet50 in Table \ref{table_pe_types_accuracy}, we observe that the ideal scenario for the best performance per unit area and energy efficiency occurs when all layers are quantized to 2-bit (LPA-2), albeit with poor accuracy. Conversely, the best quantization performance is achieved when all layers are quantized to 8-bit (LPA-8), but with lower performance per unit area and energy efficiency. Despite incorporating mixed-precision support, LPA-2/4/8 achieves accuracy tending to the ideal scenario for both metrics, demonstrating a balanced trade-off.

\section{Conclusion}
This paper presents an algorithm-hardware co-design featuring a novel composite data type, LP, which combines posits and LNS. LP dynamically adapts to diverse DNN parameter distributions and dynamic ranges by configuring bit fields. LPQ, an automated quantization framework employing genetic algorithms optimizes LP parameters through a global-local contrastive objective. We also propose LPA that integrates a unified LP PE in a systolic array architecture. Our co-design achieves on average <1\% accuracy drop and significantly improves PPA and energy efficiency compared to SoTA quantization accelerators and frameworks.

\section{Acknowledgements}
This work was supported in part by CoCoSys, one of seven centers in JUMP 2.0, a Semiconductor Research Corporation (SRC) program sponsored by DARPA.

\bibliographystyle{ACM-Reference-Format}
\footnotesize{%
\bibliography{main}
}
\end{document}